\documentclass[12pt]{article}        % For preprint
\usepackage{jheppub} % for details on the use of the package, please
\usepackage{amssymb,amsfonts,bm,dsfont,mathtools}
\usepackage{epsfig}
\def\be{\begin{eqnarray}}
\def\ee{\end{eqnarray}}
\newcommand{\nn}{\nonumber}
\newcommand\para{\paragraph{}}
\newcommand{\ft}[2]{{\textstyle\frac{#1}{#2}}}
\newcommand{\eqn}[1]{(\ref{#1})}

\def\Dslash{\,\,{\raise.15ex\hbox{/}\mkern-12mu D}}
\def\Dbarslash{\,\,{\raise.15ex\hbox{/}\mkern-12mu {\bar D}}}
\def\delslash{\,\,{\raise.15ex\hbox{/}\mkern-9mu \partial}}
\def\delbarslash{\,\,{\raise.15ex\hbox{/}\mkern-9mu {\bar\partial}}}
\def\pslash{\,\,{\raise.15ex\hbox{/}\mkern-9mu p}}
\def\calDslash{\,\,{\raise.15ex\hbox{/}\mkern-12mu {\cal D}}}

\newcommand{\Z}{{\mathbb Z}}

\newcommand{\R}{{\mathbb R}}
\newcommand{\C}{{\mathbb C}}

\newcommand{\bx}{{\bf x}}

\def\lae{\mathrel{\mathop{\smash{\lower .5 ex \hbox{$\stackrel<\sim$}}}}}
\def\lae{\mathrel{\mathop{\smash{\lower .5 ex \hbox{$\stackrel>\sim$}}}}}

\newcommand{\x}{\mathbf{x}}
\newcommand{\p}{\mathbf{p}}

\renewcommand{\r}{\mathbf{r}}
\newcommand{\z}{\mathbf{z}}
\newcommand{\0}{\mathbf{0}}
\newcommand{\X}{\mathbf{X}}
\renewcommand{\P}{\mathbf{P}}
\newcommand{\J}{\mathbf{J}}
\renewcommand{\H}{\mathcal{H}}
\newcommand{\M}{\mathcal{M}}

\newcommand{\y}{\mathbf{y}}
\newcommand{\br}{{\bf r}}

\newcommand{\A}{\mathbf{A}}

\newcommand{\wM}{\widehat{\mathcal{M}}}

\newcommand{\ket}[1]{|#1\rangle}

\title{On the Hilbert Space of Dyons}

\author{Rishi Mouland and David Tong}

\affiliation{Department of Applied Mathematics and Theoretical Physics \\ University Cambridge, CB3 0WA, UK}

\emailAdd{r.mouland, d.tong@damtp.cam.ac.uk}

\abstract{
We revisit the construction of the Hilbert space of non-relativistic particles moving in three spatial dimensions. This is given by the space of sections of a line bundle that can in general be topologically non-trivial. Such bundles are classified by a set of integers---one for each pair of particles---and arise physically when we describe the interactions of dyons, particles which carry both electric and magnetic charges. The choice of bundle fixes the representation of the Euclidean group carried by the Hilbert space. These representations are shown to recover the `pairwise helicity' formalism recently discussed in the literature. 
}

\begin{document} 

\maketitle
\flushbottom

\section{Introduction}

In this paper, we return to an old and very basic question:  what is the Hilbert space of $n$ particles, each moving on  $\mathbb{R}^3$? Naively, this seems like the kind of question that was answered in our first course on quantum mechanics. However, as is well known, there can be a twist.
 
\para
The twist comes because wavefunctions in quantum mechanics are not functions on configuration space. Instead, they are sections of line bundles over configuration space. And, for particles moving in $\R^3$, these line bundles can have interesting topology. It turns out that these topologically non-trivial bundles arise when the particles in question carry both electric and magnetic charges. In other words, they arise for dyons.

\para
Our interest in this problem was motivated by a series of papers by Csaki et. al. \cite{csaba1,csaba2}, following earlier work of Zwanziger \cite{zwan,zwan0}. They argued that the usual description of multi-particle Fock spaces as the tensor product of single particle Hilbert spaces needs some refinement when the particles in question are dyons. This refinement comes in the form of what these authors called ``pairwise helicity", an extra phase that  dyonic particles pick up under a rotation $R\in SU(2)$. Specifically, the rotation acts on elements of the Fock space, with elements labelled by the momenta ${\bf p}_i$, as 
\be \ket{{\bf p}_1,\ldots,{\bf p}_n} \rightarrow  e^{-i\varphi} \ket{{\bf p}'_1,\ldots,{\bf p}'_n} \ ,\label{pairwise}\ee
where ${\bf p}_i'=R{\bf p}_i$. There is a great deal of redundancy in the phase $\varphi$ because there's no canonical way to compare the phases of states that correspond to different rays in the Hilbert space. Indeed, the sole invariant information arises when the $\p_i$ are colinear and we consider a rotation with $\p'_i=\p_i$: then the phase $\varphi$ is physical. 

\para
In general, the phase $e^{-i\varphi}$ is a product over pairs of particles that takes the form
\be \varphi = \frac{1}{4\pi}\sum_{i<j}^n(e_ig_j-e_jg_i) \varphi_{ij}\ .\ee
Here $\varphi_{ij}$ is a somewhat cumbersome (and largely gauge dependent!) function of the rotation $R$ and the direction of ${\bf p}_i-{\bf p}_j$. Importantly, the proportionality factor is the duality invariant product of the electric charges $e_i$ and magnetic charges $g_i$ of the two particles.

\para
For $n=2$ particles, the existence of the  phase follows in a reasonably straightforward way by looking at the profiles of the electric and magnetic fields and seeing  how they transform under the little group that leaves the relative momentum ${\bf p}_1-{\bf p}_2$ unchanged. 
This is seen most clearly in the work of Wu and Yang \cite{wuyang} which we review later.  For $n\geq 3$ particles, however, the calculation is less direct. It was intuited by Zwanziger as an obvious generalisation of the two-particle case, and subsequently derived by studying a kind of doubled field theory in which both electric and magnetic potentials are introduced before projection onto a particular duality frame \cite{zwan0}. (See, also, \cite{brown1,brown2,yale} for further commentary on these ideas.)

\para
The purpose of this paper is to provide a clean derivation of the pairwise helicity formula \eqn{pairwise}. We do this by looking not at the Fock space of relativistic field theories, but instead at non-relativistic quantum mechanics. We will show that the additional phase is a direct consequence of topology. This arises because the wavefunctions that describe dyons are not functions, but sections of a line bundle that may be topologically non-trivial.  We start just with a collection of $n$ non-relativistic particles and show that the formula \eqn{pairwise} naturally arises as the most general transformation  under a rotation.  In addition, we will also derive an expression  for a lower bound on the relative angular momentum carried by a collection of dyons.

%Our goal here is in part to explain the connection of this earlier work to the more recent ideas of pairwise helicity and to link it to results in the mathematics literature. We also extend the earlier results of \cite{sorkin} in  number of ways, including understanding how things transform under rotations

\section{The Classical Theory}

We start by describing a simple, classical set-up. We will be interested in the dynamics of $n$ non-relativistic particles, each moving in $\R^3$. Crucially, we impose one condition: the particles cannot sit on top of each other. 

\para
This means that the configuration space is not $\mathbb{R}^{3n}$, as one might naively think,  but is instead what mathematicians refer to as {\it the} configuration space ${\rm Conf}_n(\mathbb{R}^3)$ and what we will denote as ${\cal M}_n$. If each particle has position $\bx_i\in\R^3$,  with $i=1,\ldots,n$, then the configuration space is spanned by the coordinates
\be
  \M_n = \Big\{ \bx_i\ :\ \bx_i\neq \bx_j\ {\rm for}\ i\neq j\Big\}\ .\ee
Importantly, coincident points are excised. 
 
\para
Our particles  will interact with some fixed, background $U(1)$ connection. This requires some explanation because it  is deceptively familiar.  In  electromagnetism, we're used to thinking of the  magnetic field as a background connection over space $\R^3$. Each  particle then experiences the same Lorentz force law, with the strength determined by the particle's electric charge. Here, instead, we will consider connections over the configuration space $\M_n$. It's not immediately obvious how to interpret this physically, but we will soon learn how these connections  relate to things we know from electromagnetism.

\para
 It is at this point that we find the novelty that underlies much of this paper: a $U(1)$  principal bundle over ${\cal M}_n$ can be topologically non-trivial. The classification of such bundles was found in \cite{cohen} (see also \cite{dmann}), following \cite{arnold,cohen2}, and is given by
\be
  H^2(\M_n,\mathbb{Z}) = \mathbb{Z}^{ n(n-1)/2}\ .
  \label{Doebner and Mann result}
\ee
The fact that we need  ${n\choose 2} = \ft12 n(n-1)$ integers to specify a bundle is interesting. It is highly suggestive that these integers can be associated to pairs of particles. Later, when we construct these bundles explicitly, we will see that this is indeed the case and, in doing so, re-derive \eqn{Doebner and Mann result}. For now, we take this as given and write the $\ft12 n(n-1)$ integers as
\be w_{ij} = -w_{ji}\ \ \ {\rm with}\ \ \ i,j=1,\ldots, n\ .\ee
The first question that we would like to answer is: what is the physical interpretation of these integers?

\subsection{The Classical Action}\label{actionsec}

To understand the physics, it's useful to first write down a classical action describing the motion of the $n$ particles. We introduce the coordinates $x^a$ on $\M_n$, and the metric $g_{ab}$, with $a,b=1,\ldots,3n$. If we introduce the $U(1)$ gauge field $A_a$, then it's tempting to write the action
\be S = \int dt\ \left(\frac{1}{2}g_{ab} \dot{x}^a\dot{x}^b  + k A_a(x) \dot{x}^a\right)\ ,\label{naive}\ee
with $k$ a parameter that is something akin to electric charge. This is fine when the $U(1)$ bundle is topologically trivial, or when the action is restricted to a given patch of a topologically non-trivial bundle. For many issues, this will be sufficient. However, there will be times when we want to be more careful and work with an action that holds everywhere on $\M_n$, even when $w_{ij}\neq 0$. For this,  we need a minor modification. 

\para
We should really think of the target space of our theory as a $U(1)$ principal bundle,
\begin{align}
  U(1) \hookrightarrow P \xrightarrow{\pi} \M_n \ .
\end{align}
We will shortly see that such bundles are parameterised by the integers $w_{ij}$. We also choose a connection $\omega$, which is a 1-form on $P$. In some patch with coordinates $(x^a, e^{i\sigma})$, it takes the form
\begin{align}
  \omega = d\sigma + A \ ,
\end{align}
where $A=A_a(x) dx^a$ is the locally defined 1-form on $\M_n$ that appeared in our action \eqn{naive}. The key point is that while $A$ is defined only locally, the 1-form $\omega$ is globally defined. This allows us to write down an action that holds everywhere.

\para 
This action should be thought of as a function of a curve $\bar{\gamma}$ in $P$. Let $\gamma = \pi \circ \bar{\gamma}$ be the projection of this curve down to $\M_n$, and let $\dot{x}^a$ be the tangent to this curve with respect to a parameter $t$. Then a better version of \eqn{naive} is
\be
  S[\bar{\gamma}] = \int_\gamma dt\ \frac{1}{2}g_{ab}(x) \dot{x}^a \dot{x}^b + k\int_{\bar{\gamma}} \omega \ .
  \label{proper action}
\ee
This action was previously considered in some detail by Friedman and Sorkin \cite{sorkin}. If $\gamma$ lies in some patch, then we have
\be  S[\bar{\gamma}] = \int_\gamma dt\ \left(\frac{1}{2}g_{ab}(x) \dot{x}^a \dot{x}^b + k(\dot{\sigma}+A_a \dot{x}^a)
\right) \ . \label{better}\ee
If we drop the total derivative $\dot{\sigma}$ then we get back to our previous, naive action \eqn{naive} and it looks like this was a rather pedantic, pointless excursion. But, as we will see later, there will be occasions when we want to move between patches when the action \eqn{better} will prove useful. 

\para
There is a second, closely related, classical action which describes motion on the so-called {\it bundle metric}. We describe this in Appendix \ref{bundlesec}. 

\subsection{Topology and Dyons}

Now we can address our question: what is the physical meaning of the integers $w_{ij}$ that characterise the topology of the bundle? This is simplest to see if we first look at two particles for which the configuration space takes the form
\be \M_2 = \R^3\times (\R^3 \setminus \{{\bf 0}\})\ .\ee
Here the first factor describes the centre of mass, while the second  factor is parameterised by the relative position, ${\bf r}=\bx_1-\bx_2$. The interesting topology sits in the second factor and arises because we have excised the origin. 

\para
For our immediate purposes, we can get away with the cheap version of the action (\ref{naive}). Ignoring the centre of mass motion, the relative motion is described by the action,
\be S_2 = \int dt \left(\frac{1}{2}\dot{\br}\cdot\dot{\br} + \dot{\br}\cdot{\bf A}(\br)\right)\ . \ee
We have set the parameter $k=1$ in \eqn{naive}. (It doesn't really play much of a role in the story; just rescaling other charges by a constant amount. We will later reinstate it when we come to the quantum theory.) The topology in the connection means that the associated magnetic field ${\bf B} = \nabla \times {\bf A}$ has first Chern class
\be \frac{1}{2\pi} \int_{{\bf S}^2_\infty} d{\bf S}\cdot {\bf B} = w \in \Z\ .\ee
This, of course, is a magnetic monopole. But it's a monopole in the relative position, which means that one particle is electrically charged and the other magnetically charged. We have no way to determine which is which based on the information above; it is just the relative charges that are captured by the topology. As shown by Friedman and Sorkin \cite{sorkin} by studying the equations of motion arising from the action, this relative charge is identified as the duality invariant combination 
\be w = \frac{1}{4\pi} (e_1g_2-g_1e_2)\ \in \Z \ ,\ee
where $e_i$ and $g_i$ are the electric and magnetic charges of the $i^{\rm th}$ particle. 

\para
What about the case of general $n$? Once again, the equations of motion make evident that the bundle characterised by the winding numbers $w_{ij}$ describes dyons with relative mutual charges
\be w_{ij} = \frac{1}{4\pi} (e_ig_j-g_ie_j)\ .\label{zwany}\ee
However, this can't be the full story: there are only $2n$ electric and magnetic charges, while there are $\ft12 n(n-1)$ different winding numbers $w_{ij}$. For example, with $n=4$ particles, the condition \eqn{zwany} implies that $w_{12} w_{34} - w_{13} w_{24} + w_{14} w_{23} = 0$. But it is certainly possible to construct sensible bundles that do not obey this constraint (and we will do so later). This means that the action \eqn{naive} must describe something more  than particles carrying an electric and magnetic charge.

\para
In fact, the right physics does not need that much more. In general, the action \eqn{naive} (or its improved version \eqn{better}) describes particles  with magnetic and electric charges under a $U(1)^r$. The winding numbers have the interpretation
\be  w_{ij} = \frac{1}{4\pi}(\vec{e}_i\cdot\vec{g}_j - \vec{g}_i\cdot\vec{e}_j)\ , \ee
where each $\vec{e}_i$ and $\vec{g}_i$ is now an $r$-vector. There is always some choice of $r$ for which such vectors can be found for any choice of $w_{ij}$. (On general grounds, one expects that $r$ should scale as $n$.)

\para
There is something a little surprising about this. We started by introducing a single $U(1)$ background gauge field over $\M_n$, and yet find that it describes the Lorentz force law between dyons carrying charges in some larger $U(1)^r$.  

\para
Note also that the winding numbers $w_{ij}$ are duality invariant and the $U(1)$ connection over $\M_n$ should not be thought of as magnetic, or electric: it too is duality invariant. But the action \eqn{naive} includes only the Lorentz force law, and not the Coulomb repulsion which must be put in separately. That is not surprising: the Lorentz force law and Coulomb law are related only by relativistic physics and we are firmly in the non-relativistic realm. If we were to add a Coulomb term, we would break the duality symmetry and specify the individual charges $\vec{e}_i$ and $\vec{g}_i$, rather than just the combinations $w_{ij}$.

\subsection{Symmetries}\label{rotationsec}

We would like to understand how symmetries of our theory act, with particular focus on rotations. The base configuration space ${\cal M}_n$ admits an $SO(3)$ rotational symmetry, generated by the vector fields
\begin{align}
  \xi^\alpha = -i \sum_i \epsilon^{\alpha \beta\gamma}\x_i^\beta \frac{\partial}{\partial \x_i^\gamma}\ .
\end{align}
 We then ask: when and how does this give rise to a symmetry of the theory defined by the action (\ref{proper action})? This is the kind of question previously addressed by Jackiw and Manton \cite{jackiw}. The key takeaway will be that as we rotate in the base, we must also move in the fibre.

\para 
The first task is to lift the $SO(3)$ action to an action on the full bundle $P$. This will generically result in  an action of the double cover $SU(2)$. A powerful result of Palais and Stewart \cite{stewart,palais} ensures that such a lift exists, and furthermore that it is unique up to conjugation by a gauge transformation.

\para
The first term in our action (\ref{proper action}) is then invariant under this lifted group action, so long as the metric is rotationally-symmetric,
\begin{align}
  \mathcal{L}_{\xi^\alpha}g=0\ .
\end{align}
Indeed, if our particles move on flat $\R^3$, then the configuration space $\M_n$ inherits a natural rotationally invariant metric. That leaves us with the second term in \eqn{proper action}, involving the connection on the bundle. We require that the $SU(2)$ action preserves our  chosen connection; this can be  ensured if and only if the field strength $F=dA$, which is a globally defined on $\M_n$, is rotationally-symmetric,
\be \mathcal{L}_{\xi^\alpha}F =0 \ .\nn\ee
In this case, the lift of the $SU(2)$ action to the full bundle is unique and is generated by the vector fields
\begin{align}
   \tilde{\xi}^\alpha = \xi^\alpha_H - \frac{i}{2} \epsilon^{\alpha\beta\gamma}F(\xi^\beta,\xi^\gamma)\eta\ .
   \label{lifted vector fields}
\end{align}
There is some notation to unpack here. The vector field $\eta$ on $P$ is locally given by $\eta = \frac{\partial}{\partial \sigma}$, meaning that it moves us along the fibre of $P$. It is normalised so that $\omega(\eta)=1$. Meanwhile $\xi_H$ is the horizontal lift of the original generators $\xi$, which locally takes the form
\be \xi_H = \xi^a\partial_a - (\xi^aA_a)\eta\ .\ee
This  horizontal lift is the naive way of lifting a vector field from the base the full bundle; we will later see that this gives rise to the usual covariant derivative when we quantise. What's more interesting is the extra term in \eqn{lifted vector fields} which, in coordinates, reads $F(\xi^\beta,\xi^\gamma) = F_{ab}\xi^{\beta\,a}\xi^{\gamma\,b}$. This is what will ultimately result in the additional phase in the wavefunction \eqn{pairwise}.

%Let us unpack some of the notation here. $\eta$ is the vertical vector field on $P$ normalised such that $\omega(\eta)=1$, and so $\eta=\partial_\sigma$ locally. Given a vector field $v$ on $\M_n$, we denote by $v_H$ its horizontal lift with respect to the connection $\omega$; in some local patch,  $v_H = v^a (\partial/\partial x^a) - (v^a A_a)\eta$. Meanwhile, for vector fields $v,w$ on $\M_n$, we have as usual $F(w,v) = F_{ab}w^a v^b$. 

\para
It is straightforward to check that
\begin{align}
  \mathcal{L}_{\tilde{\xi}^\alpha}\omega =0\ ,
\end{align}
as required, and that
\begin{align}
  [\tilde{\xi}^\alpha,\tilde{\xi}^\beta]			&= i \epsilon^{\alpha\beta\gamma}\tilde{\xi}^\gamma 	\ .
  \label{symmetry algebra}
\end{align}
The additional term, involving the field strength $F$ is \eqn{lifted vector fields} is needed for both results to hold.

\subsection{Constructing the Bundles by Pullback}

We would next like to construct the principal $U(1)$ bundles over  $\M_n$. In doing so, we will provide an alternative way to classify the bundles, providing an independent derivation  of the result (\ref{Doebner and Mann result}). 

\para
The case of two particles is straightforward. The base $\M_2$ takes the simple form
\begin{align}
  \M_2 = \mathbb{R}^3 \times (\mathbb{R}^3\setminus \{\0\}) = \mathbb{R}^3 \times \mathbb{R}^+ \times S^2\ .
  \label{M2 factorisation}
\end{align}
The $U(1)$ principal bundles $P_2[w]$ over $\M_2$ are well-known  and are classified by a single integer $w_{12}=w\in \Z$. Roughly speaking, this counts the number of times that the $U(1) \cong S^1$ winds as we move around $S^2$. More precisely, up to an isomorphism these bundles always take the form
\begin{align}
  P_2[w] = \mathbb{R}^3 \times \mathbb{R}^+\times B[w]\ .
  \label{P2 decomp}
\end{align}
Here $B[w]$ is the bundle
\begin{align}
  U(1)\hookrightarrow B[w] \to S^2\ ,
\end{align}
with first Chern class $w\in H^2(S^2,\Z) = \Z$. This integer completely specifies the bundle. In particular, the $w=0$ bundle is trivial, $B[0]\cong S^1\times S^2$, while otherwise we have the lens space $B[w] \cong S^3/\mathbb{Z}_{|w|}$ with orientation fixed by the sign of $w$.

\para
The situation for $n\ge 3$ particles is less tractable. The issue is that the base $\M_n$ does not admit a nice factorisation of the form (\ref{M2 factorisation}) for $n\ge 3$, and so we can't arrive so immediately at the possible $U(1)$ bundles over it.

\para
At this stage, we use a trick. This is the trick that will ultimately allow us to arrive at the key result \eqn{pairwise} for pairwise helicity. 
The idea is to work not with the configuration space ${\cal M}_n$, but instead on a bigger space $\widehat{\M}_n$ defined as
\begin{align}
  \widehat{\M}_n = \R^3 \times \left(\R^3\setminus\{\0\}\right)^{n(n-1)/2}\ .\label{widehatm}
\end{align}
The coordinates on this bigger space are the centre of mass ${\bf X}$, together with the separations ${\bf r}_{ij}$, with $i,j=1,\ldots,n$ and $i<j$, between any pair of particles. Importantly, however, we don't impose any constraint on these separations so, for example, ${\bf r}_{12}$ and ${\bf r}_{23}$ and ${\bf r}_{13}$ are all independent coordinates on $\widehat{\M}_n$, while they are clearly related in the original configuration space $\M_n$.

\para
There is a natural embedding of $\M_n$ into $\widehat{\M}_n$, defined simply by mapping a point with coordinates $\bx_i$ in $\M_n$ to the point with coordinates  ${\bf X}=\frac{1}{n}\sum_i\bx_i$  and ${\bf r}_{ij} = \bx_i-\bx_j$ in $\widehat{\M}_n$. We write this map as 
\begin{align}
  \phi:\M_n \mapsto \widehat{\M}_n\ .
\end{align}
%
%
%\begin{align}
%  \phi(\x_1,\dots,\x_n) = \left(\X=\tfrac{1}{n}(\x_1+\dots+\x_n), \r_{ij} = \x_i - \x_j\right)
%  \label{phi map}
%\end{align}
%
Any $U(1)$ bundle over $\widehat{\M}_n$ can be pulled back by $\phi$ to define a bundle over $\M_n$. Furthermore, a connection on this bigger bundle pulls back to a connection on the bundle over $\M_n$.

\para 
Because $\widehat{\M}_n$ has a simple factorisation \eqn{widehatm}, the bundles  $\widehat{P}_n[w]$ over $\widehat{\M}_n$ are easy to classify; they take the form
\begin{align}
  \widehat{P}_n[w] = \R^3 \times \left(\bigtimes_{i<j} \Big(\R_{>0} \times B[w_{ij}]\Big)\right)\ .
\end{align}
These bundles are fully classified by a set of integers $w_{ij}=-w_{ji}$, since $H^2(\widehat{\M}_n,\Z) = \Z^{n(n-1)/2}$. Given a connection, each integer specifies the flux of $F$ through a 2-cycle that wraps the excised origin in the $\r_{ij}$ plane.

\para 
We then  define the bundle $U(1)\hookrightarrow P_n[w]\to \M_n$ over our original configuration space  by the pullback $P_n[w]=\phi^* \widehat{P}_n[w]$. For two particles, we just have $P_2[w]\cong \widehat{P}_2[w]$, and we trivially recover the above construction.

\para 
All we have done thus far is provided a tool by which we can explicitly construct $U(1)$ bundles over $\M_n$.  But that's a far cry from our initial claim that this construction allows us to classify all such bundles. There are two  things that could go wrong. First, even though the bundles $\widehat{P}_n[w]$ are topologically inequivalent, we do not know whether their pullbacks $P_n[w] = \phi^* \widehat{P}_n[w]$ are topologically inequivalent.  Secondly, there could be further topologically inequivalent bundles over $\M_n$ which do not arise as the pullback of some bundle over $\widehat{M}_n$. 

\para
It turns out that neither of these issues actually arise: the isomorphism classes of bundles over $\M_n$ are precisely labelled by integers $w_{ij}=-w_{ji}$, with the bundles $P_n[w]=\phi^* \widehat{P}_n[w]$ providing a full set of topologically-inequivalent representatives. We relegate a proof of this fact to Appendix \ref{multi-particle bundle appendix}, thus providing an independent derivation of (\ref{Doebner and Mann result}).

\subsubsection*{Symmetric Connections}

This pullback construction also provides a way to write down symmetric connections on the bundle $P[w]$ over configuration space. 
Again, the case of two particles is straightforward. We have a connection of uniform curvature on the bundle $B[w]$ with  $\frac{1}{2\pi}\int_{S^2}F = w$. This is a familiar magnetic monopole field strength, now in the space parameterised by the relative position of the two particles, 
\begin{align}
  F= \frac{w}{4} \frac{\epsilon^{\alpha\beta\gamma}\x_{12}^\gamma }{|\x_{12}|^3} d\x_{12}^\alpha \wedge d\x_{12}^\beta \ ,
  \label{two particle curvature}
\end{align}
where $\x_{12}=\x_1-\x_2$.  Wu and Yang \cite{wuyangfirstpaper} showed that there is a two-patch open covering of $P_2[w]$ such that the gauge field in each patch takes the familiar form,
\begin{align}
  A_N = -\frac{1}{2}w(\cos\theta-1)d\phi\ ,\qquad A_S = -\frac{1}{2}w(\cos\theta+1)d\phi\ ,
  \label{wu-yang gauge field}
\end{align}
where $(\theta,\phi)$ are spherical polar coordinates associated to the 3-vector $\x_{12}$.

\para
Once again using the bigger bundles $\widehat{P}_n[w]$, this construction generalises immediately to multiple particles. We first fix the a curvature on $\widehat{P}_n[w]$, which takes the form 
\be \hat{F}=\sum_{i<j}F^{(ij)}\ ,\ee
where $\hat{F}^{(ij)}$ is a curvature in $B[w_{ij}]$ with $\frac{1}{2\pi}\int_{S_{(ij)}^2}\hat{F}^{(ij)} = w_{ij}$. Given such a connection $\hat{\omega}$ on $\widehat{P}_n$, we can then simply pullback  to define a connection $\omega=\phi^*\hat{\omega}$ on $P_n[w]$, whose curvature $F$ by construction has the desired symmetry properties. Explicitly, we have
\begin{align}
  F = \frac{1}{4} \epsilon^{\alpha\beta\gamma} \sum_{i<j} \frac{w_{ij}\x_{ij}^\gamma }{|\x_{ij}|^3} d\x_{ij}^\alpha \wedge d\x_{ij}^\beta \ ,
\end{align}
where, since we have pulled back to $P_n[w]$, the $\x_{ij}$ are not independent functions on $\M_n$ when $n\ge 3$. The gauge field can be written in a form analogous to (\ref{wu-yang gauge field}) in the various patches on $P_n[w]$.

\para
With this connection in hand, we can be a little more explicit in the expression (\ref{lifted vector fields}). The all-important additional term in the generator for rotational symmetries is proportional to
\begin{align}
  \epsilon^{\alpha\beta\gamma}F(\xi^\beta,\xi^\gamma) = -\sum_{i<j}w_{ij}\frac{\x _{ij}^\alpha}{|\x_{ij}|}\ .
\end{align}
This, ultimately, will translate into the phase picked up by the wavefunction under rotations, as we now show.

\section{The Quantum Theory}

We now turn to the quantum theory. Our goal is to construct the Hilbert space of the theory and determine the representation of the Euclidean group that it carries.

\subsection{Basic Quantisation}

We work with quantum mechanics on the bundle $P_n[w]$ where, recall, the winding of the $U(1)$ fibre over the base configuration space $\M_n$ is specified by the $\ft12 n(n-1)$ integers $w_{ij}$. The action is given by \eqn{better} 
\be  S[\bar{\gamma}] = \int_\gamma dt\ \left(\frac{1}{2}g_{ab}(x) \dot{x}^a \dot{x}^b + k(\dot{\sigma}+A_a \dot{x}^a)
\right) \ . \label{better2}\ee
Upon quantisation, the constraint $\pi_\sigma = \partial L/\partial \dot{\sigma} = k$ imposes a physical state condition. The physical Hilbert space $\H_n[w]$ is thus identified as the space of sections of a complex line bundle of charge $k$ associated to $P_n[w]$. Our connection on $P_n[w]$ then induces a covariant derivative on $\H_n[w]$.

\para 
For two particles, the Hilbert space $\H_2[w]$ is straightforward. By the decomposition (\ref{P2 decomp}), we can write
\begin{align}
  \H_2[w] = \H_\text{com}\otimes \H_\text{rel}[w]\ .
\end{align}
The first factor $\H_\text{com}$ is just a Hilbert space of functions of the centre-of-mass position $\X=\frac{1}{2}(\x_1+\x_2)$. If we want to consider momentum eigenstates, we should impose plane-wave normalisability. The second factor $\H_\text{rel}[w]$ is the space of sections of a complex line bundle of first Chern class $kw$ over $\R^3\setminus \{\0\}$, spanned by the relative position $\x_{12}=\x_1-\x_2$. Choosing the usual north-south patchwise covering, on the equatorial overlap a state $\psi_\text{rel}\in \H_\text{rel}[w]$ satisfies
\begin{align}
  \psi_\text{rel}^\text{N}(\x_{12}) = e^{ikw\phi}\psi_\text{rel}^\text{S} (\x_{12})\ ,
  \label{wavefunction patching}
\end{align}
where $\phi$ is the azimuthal part of $\x_{12}$. A basis for $\H_\text{rel}[w]$ is provided by monopole spherical harmonics \cite{wuyang}, which furnish representations of $SU(2)$ of spins $j=\frac{1}{2}|kw|,\frac{1}{2}|kw|+1,\dots$.

\para
For $n>2$ particles, we can proceed using the construction of $P_n[w]$ as a pullback of the bigger bundle $\widehat{P}_n[w]$. What this means is that $\H_n[w]$ is a linear subspace of a bigger Hilbert space,
\begin{align}
  \widehat{\H}_n[w] = \H_\text{com}\otimes \left(\bigotimes_{i<j} \H_\text{rel}[w_{ij}]\right)\ .
  \label{bigger Hilbert space}
\end{align}
The pullback $\phi^*$ induces a linear map on $\widehat{\H}_n[w]$, whose image is the Hilbert space we want, $\H_n[w]$.

\subsection{It's Just a Phase We're Going Through}

The action (\ref{lifted vector fields}) of $SU(2)$ on $P_n[w]$ induces a representation of $SU(2)$ on $\H_n[w]$. It is generated by the operators $\J = (J_1,J_2,J_3)$, which locally take the form
\begin{align}
  \J = -i \sum_i \x_i \times \left( \frac{\partial}{\partial \x_i} - ik \A_i\right) - \frac{k}{2}\sum_{i<j} w_{ij} \frac{\x_{ij}}{|\x_{ij}|}\ ,
  \label{rotation operators}
\end{align}
where $A=\A_i(x) \cdot d\x_i$. We also have the total linear momentum operator
\begin{align}
  \P = - i \sum_i \left( \frac{\partial}{\partial \x_i} - ik \A_i\right) = - i \sum_i\frac{\partial}{\partial \x_i}\ ,
  \label{linear momentum}
\end{align}
acting only on the centre-of-mass coordinate. Then, $(\P,\J)$ generate the Euclidean group $\R^3 \rtimes SU(2)$ of symmetries\footnote{It is easy to show that for our choice of connection, the action (\ref{proper action}) admits a symmetry under the full Euclidean group, which arises as a lift of the action of the Euclidean group on $\M_n$ to an action of $\R^3 \rtimes SU(2)$ on $P_n[w]$. The additional translation generators give rise to the $\P$ stated here.} of the theory.

\para
There are  some immediate consequences of (\ref{rotation operators}) and (\ref{linear momentum}). Consider a collection of position eigenstates\footnote{Concretely, in any local patch containing the point $(\x_1',\dots,\x_n')$,  $\ket{\x_1',\dots,\x_n'}$ is described by the wavefunction $\psi(\x_1,\dots,\x_n)=c\,\delta^{(3)}(\x_1 - \x_1')\dots\delta^{(3)}(\x_n - \x_n')$ for some $c\in \C$, which we normalise as $|c|^2=1$.} $\ket{\x_1,\dots,\x_n}\in \H_n[w]$, from which we can build any state. Under some generic Euclidean transformation $g\in \R^3 \rtimes SU(2)$, we have
\begin{align}
  g\ket{\x_1,\dots,\x_n} \propto \ket{g\x_1,\dots,g \x_n}\ .
  \label{generic Euclidean group transf}
\end{align}
The constant of proportionality is some phase. Whenever $g\x_i \neq \x_i$ for some $i$, the state $\ket{g\x_1,\dots,g\x_n}$ is associated to a different ray in the Hilbert space from $\ket{\x_1,\dots,\x_n}$ and there is no physical meaning to the phase: it can always be absorbed by an appropriate a gauge transformation, $\ket{\x_1,\dots,\x_n}\to e^{i f(\x_1,\dots,\x_n)}\ket{\x_1,\dots,\x_n}$. This is not true however for $g$ such that $g\x_i=\x_i$ for all $i$. For generic $\x_i$ this occurs only for $g=-\mathds{1}_2\in SU(2)$, under which we have
\begin{align}
  (-\mathds{1}_2)\ket{\x_1,\dots,\x_n} = (-1)^F \ket{\x_1,\dots,\x_n}\ ,
\end{align}
for some choice of `fermion number' $F\in \{0,1\}$ that we should determine. Things are more interesting when the $\x_i$ are colinear which, of course, always holds when $n=2$. Now, under a rotation $R_\phi$ by angle $\phi$ about the axis connecting the $\x_i$, we must have
\begin{align}
  R_\phi \ket{\x_1,\dots,\x_n} = e^{iq\phi/2} \ket{\x_1,\dots,\x_n}\ ,
\end{align}
for some $q$, where the global structure of the symmetry group means we must have $q\in \Z$. This phase \textit{is} physical; we can't gauge it away. Note that once we know $q$ for some colinear arrangement, by continuity we can read off $F=q$ (mod $2$) for all states. In particular, each $q$ arising from each distinct colinear arrangement is equal modulo $2$.

\para
It is important to appreciate that the integers $\{q\}$, one for each colinear arrangement, along with the fermion number $F$ that they determine, constitute the \textit{complete} gauge-invariant information contained in the action of $\P$ and $\J$. Concretely, the states in the $\R^3\rtimes SU(2)$ orbit of some $\ket{\x_1,\dots,\x_n}$ span an induced representation of $\R^3\rtimes SU(2)$, which is fixed entirely by the one-dimensional representation of $\ket{\x_1,\dots,\x_n}$ under the stabiliser subgroup $\{g:g\x_i=\x_i\}$. This subgroup is generically $\Z_2$, with this representation fixed by $F$, but is enhanced to $U(1)$ for colinear states, with representation fixed by some $q\in \Z$.

\para
On one hand, our theory was specified by the integers $w_{ij}$, and $k$. On the other, we've found that the Euclidean group representation carried by the Hilbert space $\H_n[w]$ is entirely specified by some set of integers $\{q\}$, one for each colinear arrangement of particles. The former must determine the latter. All that remains is to determine how.
\para
This follows quite straightforwardly from the explicit form (\ref{rotation operators}) of $\J$. The phase $e^{iq\phi/2}$ depends on the ordering of the $\x_i$.  To this end,  set $\x_i = \alpha_i \y + \mathbf{m}$ for some vectors $\y,\mathbf{m}$, and order $\alpha_{\sigma(1)}> \dots > \alpha_{\sigma(n)}$ for some $\sigma\in S_n$. Then, under a rotation $R_\phi$ by angle $\phi$ about the colinear axis, we have $R_\phi \x_i = \x_i$, and
\begin{align}
  R_\phi \ket{\x_1,\dots,\x_n} = \exp\left(-\frac{ik\phi }{2}\sum_{i<j}w_{\sigma(i)\sigma(j)}\right) \ket{\x_1,\dots,\x_n}\ ,
  \label{positioneigenstatesphases}
\end{align}
where, recall, $w_{ji}=-w_{ij}$. In particular, we determine the fermion number carried by the Hilbert space,
\begin{align}
  F=k\sum_{i<j} w_{ij} \quad (\text{mod } 2)\ .
  \label{fermionnumber}
\end{align}
For example, for three particles there are three different orderings, up to reversing the orientation. If the three particles are ordered along the line as $(123)$, the phase is proportional to $(w_{12} + w_{23} - w_{31})$. For ordering $(312)$, this becomes $(w_{12} - w_{23} + w_{31})$. Finally, if the ordering is $(231)$, the phase is proportional to $(-w_{12} + w_{23} + w_{31})$. 

\subsection{Momentum Eigenstates}

Our ultimate goal is to make contact with the pairwise helicity formula (\ref{pairwise}). For this, we need a set of momentum eigenstates. Concretely, we would like a set of states $\psi_{\p_1,\dots,\p_n}\in \H_n[w]$ that are eigenstates of the total momentum operator
\begin{align}
  \P \psi_{\p_1,\dots,\p_n} = (\p_1 + \dots + \p_n)\psi_{\p_1,\dots,\p_n}\ ,
  \label{total momentum}
\end{align}
and which furnish a representation of $SU(2)$,
\begin{align}
  R \psi_{\p_1,\dots,\p_n} = D_{\p_1,\dots,\p_n}(R) \psi_{\p_1',\dots,\p_n'} \quad{\rm with}\quad \p_i' = R\p_i\quad{\rm and}\quad R=e^{i \boldsymbol{\theta}\cdot \J}\ ,
  \label{momentum eigenstate transf}
\end{align}
for a phase $D_{\p_1,\dots,\p_n}(R)$ that we must determine. Note, the transformation of the individual labels $\p_i$ is \textit{not} implied by the Euclidean algebra obeyed by $\P$ and $\J$, which only ensures that the sum transforms appropriately: $(\p_1' + \dots + \p_n') = R(\p_1 + \dots + \p_n)$.

\para
In the topologically-trivial case $w_{ij}=0$, we're done: we just write down
\begin{align}
  \psi_{\p_1,\dots,\p_n}(\x_1,\dots,\x_n) = e^{i \p_1 \cdot \x_1}\dots e^{i \p_n \cdot \x_n}\ .
\end{align}
In particular, $\psi_{\p_1,\dots,\p_n}$ is a simultaneous eigenstate of the single-particle momentum operators $-i\partial /\partial \x_i$.
The situation is very different when some $w_{ij}\neq 0$. Now, the gauge covariant single-particle momentum operators
\begin{align}
  -i \left( \frac{\partial}{\partial \x_i} - ik \A_i\right)\ ,
\end{align}
do not mutually commute\footnote{One might try to deform these operators with additional terms to make them commute. However, whenever some $w_{ij}\neq 0$, no such globally-defined term exists.} due to the necessarily non-zero curvature of $A$. Thus, there is no sense in which we can view $\psi_{\p_1,\dots,\p_n}$ as being a simultaneous eigenstate of single-particle momentum operators; the $\p_i$ are just labels, not quantum numbers. Nonetheless, it is a well-posed, non-trivial problem to find states $\psi_{\p_1,\dots,\p_n}$ satisfying (\ref{total momentum}) and (\ref{momentum eigenstate transf}).

\para
We will now construct such states $\psi_{\p_1,\dots,\p_n}$. These will be sections of a non-trivial bundle and, as such, we will be forced to think about patches and all that. However, we can already glean the most important information contained in (\ref{momentum eigenstate transf}). Namely, we are in essentially the same situation as with the position eigenstates. For generic $\p_i$, it is clear that we can absorb the phase $D_{\p_1,\dots,\p_n}(R)$ by a suitable redefinition of the $\psi_{\p_1,\dots,\p_n}$, except when $R=-\mathds{1}_2\in SU(2)$ and we have $(-\mathds{1}_2)\psi_{\p_1,\dots,\p_n} = (-1)^F \psi_{\p_1,\dots,\p_n}$ with $F$ given in (\ref{fermionnumber}). However, when the $\p_i$ lie along an axis through the origin, the state picks up a physical phase under rotations about this axis\footnote{Once again, we can phrase this more precisely in representation theoretic language. The states in the orbit of a given state $\psi_{\p_1,\dots,\p_n}$ span an induced representation of $SU(2)$ which is determined entirely by the charge of $\psi_{\p_1,\dots,\p_n}$ under its stabiliser subgroup. Generically this is $\Z_2$, with representation fixed by $F$. For states with $\p_i$ lying on an axis through the origin, the stabiliser subgroup is $U(1)$ and the representation is fixed by some $q\in \Z$.}. These phases are however now a little more fiddly to determine.

\para
The key fact we need, and which we will shortly show, is that it is possible \textit{locally} to make $\J$ very simple. Let us pick some unit vector $\hat{\y}\in S^2$. Then, in a patch on $\M_n$ containing the colinear points $\{\x_{\sigma(i)\sigma(j)}=\alpha\hat{\y},\,\,\alpha>0\}_{i<j}$ for some $\sigma\in S_n$, we can choose a gauge such that
\begin{align}
  \hat{\y}\cdot \J = -i \hat{\y} \cdot \sum_i \x_i \times \frac{\partial}{\partial \x_i} - \frac{k}{2}\sum_{i<j} w_{\sigma(i)\sigma(j)}\ .
  \label{simple J}
\end{align}
Next, for $\p_i = \alpha_i\hat{\y}$ with $\alpha_{\sigma(1)}> \dots > \alpha_{\sigma(n)}$, let us define
\begin{align}
  \psi_{\p_1,\dots,\p_n}(\x_1,\dots,\x_n) = e^{i\p_1\cdot \x_1}\dots e^{i\p_n \cdot \x_n}\ ,
\end{align}
in this patch. Then, consider a rotation $R_\phi$ by angle $\phi$ about $\hat{\y}$. Using (\ref{simple J}), we find
\begin{align}
  R_\phi \psi_{\p_1,\dots,\p_n}(\x_1,\dots,\x_n) = \exp\left(-\frac{ik\phi }{2}\sum_{i<j}w_{\sigma(i)\sigma(j)}\right)\psi_{\p_1,\dots,\p_n}(\x_1,\dots,\x_n)\ ,
  \label{momentum states colinear}
\end{align}
which takes precisely the same form as (\ref{positioneigenstatesphases}) for the position eigenstates. To get a full grasp of where (\ref{simple J}) comes from, it is useful to take a step back and understand how generic wavefunctions transform under generic rotations.

\subsection{Generic Rotations and Pairwise Helicity}

To start, we will look at how states in the $n=2$ particle Hilbert space transform under generic rotations. All non-triviality lies in the relative part $\H_\text{rel}$. Given some $\psi_\text{rel}\in\H_\text{rel}[w]$, and some $R\in SU(2)$, we have
\begin{align}
  \left(R \psi_\text{rel}\right)(\x_{12}) \propto \psi_\text{rel}(R^{-1}\x_{12})\ .
\end{align}
Let's pick, without loss of generality, the unit vector $\hat{\z}=(0,0,1)$. Then, we can implicitly use the $SU(2)$ action to fix the phase of $\psi_\text{rel}(\x_{12})$ in terms of that of $\psi_\text{rel}(|\x_{12}|\hat{\z})$. This is the procedure: given any $\hat{\r}\in S^2$, we choose a canonical rotation $R_\star(\hat{\r})\in SU(2)$ such that $R_\star(\hat{\r})\hat{\z} = \hat{\r}$. Then, our gauge in a patch containing $|\x_{12}|\hat{\z}$ can be entirely fixed by requiring that states satisfy
\begin{align}
  \psi_\text{rel} (\x_{12}) = \left(U(R_\star(\hat{\x}_{12}))^{-1} \psi_\text{rel}\right)(|\x_{12}|\hat{\z})\ ,
  \label{induced rep}
\end{align}
where, here and throughout, we use the notation $\hat{\x}_{12} = \x_{12}/|\x_{12}|\in S^2$ for the direction of a 3-vector. (There is a subtlety in making this choice of gauge that we will comment on below.)

\para
Having fixed this gauge, and after putting back in the centre-of-mass part, we arrive at the transformation of a generic state $\psi\in \H_2[w]$ in the 2-particle Hilbert space under some generic $R\in SU(2)$,
\begin{align}
  (R\psi)(\x_1,\x_2) = \exp\left(-\frac{ikw}{2}\theta(R,R^{-1}\hat{\x}_{ij})\right)\psi(R^{-1}\x_1, R^{-1}\x_2)\ .
  \label{2-particle state SU(2) transf}
\end{align}
Here, $\theta(R,\hat{\r})\in [0,4\pi)$ is defined for all $R\in SU(2)$ and $\hat{\r}\in S^2$ as the unique angle such that
\begin{align}
  R_\star(R\hat{\r})^{-1} R R_\star(\hat{\r}) = \exp \left(i\theta(R,\hat{\r})T^3\right)\ ,
\end{align}
where here $T^3 = \frac{1}{2}\sigma^3$ is the relevant $SU(2)$ generator. In particular, we have
\begin{align}
  \theta(R_2,\hat{\r}) + \theta(R_1,R_2\hat{\r}) = \theta(R_1R_2,\hat{\r}) \quad \text{mod }4\pi\ ,
\end{align}
which ensures that (\ref{2-particle state SU(2) transf}) does indeed define a representation of $SU(2)$. In the special case that $R_\phi$ is a rotation by $\phi$ that preserves $\hat{\r}$, i.e. $R_\phi\hat{\r}=\hat{\r}$, we have $\theta(R_\phi,\hat{\r}) = \phi$. Another important special case is when $R$ lies in the centre of $SU(2)$, for which we have $\theta(-\mathds{1}_2,\hat{\r}) = 2\pi$.

\para
Note, there is an ambiguity in how we choose $R_\star(\hat{\r})$. But it is straightforward to show that precisely this ambiguity can always be absorbed into a gauge transformation. This is simply a manifestation of the basic principle of induced representations.

\para
Let us now construct the desired momentum eigenstates. The key observation is that we can always choose $R_\star$ to be azimuthally-symmetric, which in turn ensures that $\theta(e^{i\phi T^3},\hat{\r}) = \phi$ for any $\hat{\r}$. That is to say, under a rotation by $\phi$ about $\hat{\z}$, the wavefunction $\psi$ just picks up a phase $e^{-ikw\phi/2}$. By noting the arbitrariness of $\hat{\z}$, and expanding this statement infinitesimally, we arrive precisely at the result (\ref{simple J}) for 2 particles.

\para
With this fact in hand, let us fix for any $\p_1,\p_2$ such that $\p_1-\p_2=\alpha \hat{\z}$ with $\alpha\ge 0$,
 \begin{align}
  \psi_{\p_1,\p_2}(\x_1,\x_2) = e^{i\p_1\cdot \x_1}e^{i\p_2\cdot \x_2}\ ,
\end{align}
in our patch. These states are then eigenstates of $J_3$,
\begin{align}
  J_3 \psi_{\p_1,\p_2} = -\frac{kw}{2}\psi_{\p_1,\p_2}\ .
\end{align}
We can then simply define $\psi_{\p_1,\p_2}$ at generic $\p_1,\p_2$ by a transformation of these reference states by an appropriate canonical rotation $R_\star$. When the dust settles, we end up with the states
\begin{align}
  \psi_{\p_1,\p_2}(\x_1,\x_2) = e^{i\p_1\cdot \x_1} e^{i\p_2\cdot \x_2}  \exp\left(\frac{ikw}{2}\theta(R_\star(\hat{\p}_{12})^{-1},\hat{\x}_{12})\right)\ .
  \label{2 particle momentum state}
\end{align}
As required, they satisfy
\begin{align}
  \P \psi_{\p_1,\p_2}(\x_1,\x_2) = (\p_1+\p_2)\psi_{\p_1,\p_2}(\x_1,\x_2)\ .
\end{align}
Then, under a rotation $R\in SU(2)$, they transform as
\begin{align}
  (R\psi_{\p_1,\p_2})(\x_1,\x_2) = \exp\left(-\frac{ikw}{2}\theta(R,\hat{\p}_{12})\right)\psi_{\p_1',\p_2'}(\x_1,\x_2),\quad \p_i' = R\p_i\ .
\end{align}
In particular, if $\p_i = \alpha_i \hat{\y}$ with $\alpha_1>\alpha_2$, then under a rotation $R_\phi$ by $\phi$ about $\hat{\y}$, we have
\begin{align}
  (R_\phi\psi_{\p_1,\p_2})(\x_1,\x_2) = e^{-ikw\phi/2}\psi_{\p_1,\p_2}(\x_1,\x_2)\ .
\end{align}

\subsubsection*{Global expressions}

Let us make a comment about patch-wise versus global expressions. There is a subtlety in (\ref{induced rep}) which is in fact vital. The issue is that it is not possible to choose the canonical rotation $R_\star(\hat{\r})$ smoothly over $S^2$. The best we can do is make it patch-wise smooth. For instance, covering $S^2$ with north and south patches, we find that the necessary discontinuity on the equator can be chosen precisely such that the state $\psi_\text{rel}$ is related between patches as in (\ref{wavefunction patching}), as is required for an element of $\H_\text{rel}[w]$.
 
 \para
 This discontinuity in $R_\star(\hat{\r})$ trickles down to $\theta$. Treating things more carefully with patches, one finds that for each $R\in SU(2)$, $W_R(\hat{\r})=\exp\left(\frac{i}{2}\theta(R,\hat{\r})\right)$ should be regarded as a section of a bundle $U(1)\to P \to S^2$ with first Chern class $+1$. It is precisely this fact that ensures that the formula (\ref{2-particle state SU(2) transf}) is continuous as we pass between patches. And most importantly, it allows us to reinterpret (\ref{2 particle momentum state}) and (\ref{n particle momentum state}) as global definitions of the angular momentum eigenstates, truly as sections of the relevant bundles. 

\subsubsection*{Generalisation to many particles}

We can now generalise the above results to the case of $n>2$ particles. Since $\H_n[w]$ can be found as a subspace of $\widehat{\H}_n[w]$ as in (\ref{bigger Hilbert space}), and since the latter is just built of the $\H_\text{rel}[w_{ij}]$, this is straightforward. Picking the gauge (\ref{induced rep}) in each $\H_\text{rel}[w_{ij}]$, we find
\begin{align}
  (R\psi)(\x_1,\dots,\x_n) = \exp\left(-\frac{ik}{2}\sum_{i<j}w_{ij}\theta(R,R^{-1}\hat{\x}_{ij})\right)\psi(R^{-1}\x_1,\dots R^{-1}\x_n)\ .
  \label{many state SU(2) transf}
\end{align}
Expanding this expression infinitesimally for $R=J_3$, and noting the arbitrariness of the vector $\hat{\z}$ as well as our ordering of particles, we arrive at the result (\ref{simple J}).

\para
The momentum eigenstates are then a simple generalisation of the case (\ref{2 particle momentum state}) at 2 particles,
\begin{align}
  \psi_{\p_1,\dots,\p_n}(\x_1,\dots,\x_n) = e^{i\sum_i\p_i\cdot \x_i}   \exp\left(\frac{ik}{2}\sum_{i<j}w_{ij}\theta(R_\star(\hat{\p}_{ij})^{-1},\hat{\x}_{ij})\right)\ .
  \label{n particle momentum state}
\end{align}
These satisfy $\P \psi_{\p_1,\dots,\p_n} = (\p_1+\dots + \p_n)\psi_{\p_1,\dots,\p_n}$, and under $SU(2)$ transformations tranform as
\begin{align}
  (R\psi_{\p_1,\dots,\p_n})(\x_1,\dots,\x_n) = \exp\left(-\frac{ik}{2}\sum_{i<j}w_{ij}\theta(R,\hat{\p}_{ij})\right)\psi_{\p_1',\dots,\p_n'}(\x_1,\dots,\x_n)  \ ,
  \label{mom states general rotation}
  \end{align}
where $  \p_i' = R\p_i$. 
When applied to the case of dyons, where we set $k=1$ and $w_{ij}$ is identified as in (\ref{zwany}), we recover the representation put forward in \cite{zwan,csaba1,csaba2}, based on notions of pairwise helicity.

\para
Consider then the case that the momenta are colinear, $\p_i = \alpha_i \hat{\y}_i $ with $\alpha_{\sigma(1)}> \dots >\alpha_{\sigma(n)}$. Using (\ref{mom states general rotation}), we have that under a rotation $R_\phi$ by angle $\phi$ about $\hat{\y}$,
\begin{align}
  R_\phi\psi_{\p_1,\dots,\p_n} = \exp\left(-\frac{ik\phi }{2}\sum_{i<j}w_{\sigma(i)\sigma(j)}\right)\psi_{\p_1,\dots,\p_n}\ ,
\end{align}
thus recovering (\ref{momentum states colinear}).

\subsection{Relative Angular Momentum and Spin-Statistics}

The theory (\ref{proper action}) has a yet larger symmetry: we are free to rotate the centre-of-mass coordinate $\X$ independently from the relative coordinates $\x_{ij}$. In detail, we have the splitting
\begin{align}
  \J = \J_\text{com} + \J_{\text{rel}}\ ,
\end{align}
where, when thought of as an operator on the bigger Hilbert space (\ref{bigger Hilbert space}), $\J_\text{com}$ acts only on the first factor while $\J_\text{rel}$ acts only on all the rest, and so $[\J_\text{com},\J_\text{rel}]=0$. Furthermore, both $\J_\text{com}$ and $\J_\text{rel}$ survive the restriction to $\H_n[w]$. Then, each of $\J_\text{com}$ and $\J_\text{rel}$ generate their own $SU(2)$ symmetry of the theory, of which our usual rotational symmetry is the diagonal subgroup.

\para
We can then consider organising states into irreducible representations of $SU(2)_\text{rel}$. The possible spins of such representations are $j=j_\text{min},j_\text{min}+1,\dots$. Here, $j_\text{min}\in \frac{1}{2}\Z$ is the spin of the lowest spin representation appearing in the tensor product $\otimes_{i<j}(\frac{1}{2}|kw_{ij}|)$ of $SU(2)$ representations of spins $\frac{1}{2}|kw_{ij}|\in \frac{1}{2}\Z$. In particular, $j_\text{min}$ is (half-)integer if $k\sum_{i<j}w_{ij}$ is (odd) even.

\para
We have learnt that a composite of $n$ particles carries half-integer spin whenever $k\sum_{i<j} w_{ij}$ is odd. It is natural then to ask what happens when two identical such composites interact, and in particular, are exchanged. Precisely such a scenario was considered in \cite{sorkinsecondpaper}, where it was argued that composites with half-integer spin do indeed realise fermionic statistics (see also \cite{goldhaber,csaba3}).

\subsection{A Comment on Other Dimensions}

It is interesting to note that three spatial dimensions are very special. In any other dimension, $H^2(\M_n,\mathbb{Z})$ is trivial, and so there is only one (trivial) bundle. Then, in all but two spatial dimensions, the action of the Euclidean group on the Hilbert space is essentially fixed, and is boring. In the language of \cite{csaba1,csaba2}, the only one-dimensional irreducible representation of the pairwise little group is the trivial one.

\para 
In two spatial dimensions, the non-simply-connectedness of the rotation group $U(1)$ gives us more to play with. Although there is only a single, trivial bundle, the result of \cite{palais,stewart} no longer applies, and there is a continuous family of lifts of the rotation group to a projective representation on this bundle, parameterised by an angle $\theta\in [0,2\pi)$. The resulting Hilbert space is that of a system of Abelian anyons.

\para
To see more structure---and anything non-trivial away from two and three spatial dimensions---we need to consider higher rank bundles over $\M_n$, corresponding to particles carrying their own internal degrees of freedom. It turns out that the task of classifying such bundles is significantly more involved, and indeed is an open question in general \cite{dmann}. Furthermore, even with such a bundle in hand, there may be many lifts of the symmetries of $\M_n$. It would be interesting to resolve each of these open questions, and thus derive a full classification of multi-particle Hilbert spaces.

%%%%%%%%%%%%%%%%%%%%%%%%%%%%%%%%%
%%%%%%%%%%%%%%%%%%%%%%%%%%%%%%%%%
%%%%%%%%%%%%%%%%%%%%%%%%%%%%%%%%%
%%%%%%%%%%%%%%%%%%%%%%%%%%%%%%%%%
%%%%%%%%%%%%%%%%%%%%%%%%%%%%%%%%%
%%%%%%%%%%%%%%%%%%%%%%%%%%%%%%%%%
%%%%%%%%%%%%%%%%%%%%%%%%%%%%%%%%%
%%%%%%%%%%%%%%%%%%%%%%%%%%%%%%%%%
%%%%%%%%%%%%%%%%%%%%%%%%%%%%%%%%%
%%%%%%%%%%%%%%%%%%%%%%%%%%%%%%%%%
%%%%%%%%%%%%%%%%%%%%%%%%%%%%%%%%%
%%%%%%%%%%%%%%%%%%%%%%%%%%%%%%%%%
%%%%%%%%%%%%%%%%%%%%%%%%%%%%%%%%%
%%%%%%%%%%%%%%%%%%%%%%%%%%%%%%%%%
%%%%%%%%%%%%%%%%%%%%%%%%%%%%%%%%%
\subsection*{Acknowledgements}

Thanks to Jack Smith for helpful guidance towards the classification of multi-particle bundles, and to Ofri Telem for very helpful comments on a draft of this paper. Thanks also to Nick Manton and Ben Gripaios for helpful discussions, as well as Philip Boyle Smith for comments on an earlier version. This work was supported by STFC grant ST/L000385/1, the EPSRC grant EP/V047655/1 ``Chiral Gauge Theories: From Strong Coupling to the Standard Model", and a Simons Investigator Award.  For the purpose of open access, the author has applied a Creative Commons Attribution (CC BY) licence to any Author Accepted Manuscript version arising from this submission.

%%%%%%%%%%%%%%%%%%%%%%%%%%%%%%%%%
%%%%%%%%%%%%%%%%%%%%%%%%%%%%%%%%%
%%%%%%%%%%%%%%%%%%%%%%%%%%%%%%%%%
%%%%%%%%%%%%%%%%%%%%%%%%%%%%%%%%%
%%%%%%%%%%%%%%%%%%%%%%%%%%%%%%%%%
%%%%%%%%%%%%%%%%%%%%%%%%%%%%%%%%%
%%%%%%%%%%%%%%%%%%%%%%%%%%%%%%%%%
%%%%%%%%%%%%%%%%%%%%%%%%%%%%%%%%%
%%%%%%%%%%%%%%%%%%%%%%%%%%%%%%%%%
%%%%%%%%%%%%%%%%%%%%%%%%%%%%%%%%%
%%%%%%%%%%%%%%%%%%%%%%%%%%%%%%%%%
%%%%%%%%%%%%%%%%%%%%%%%%%%%%%%%%%
%%%%%%%%%%%%%%%%%%%%%%%%%%%%%%%%%
%%%%%%%%%%%%%%%%%%%%%%%%%%%%%%%%%
%%%%%%%%%%%%%%%%%%%%%%%%%%%%%%%%%
%%%%%%%%%%%%%%%%%%%%%%%%%%%%%%%%%
%%%%%%%%%%%%%%%%%%%%%%%%%%%%%%%%%
%%%%%%%%%%%%%%%%%%%%%%%%%%%%%%%%%
%%%%%%%%%%%%%%%%%%%%%%%%%%%%%%%%%
%%%%%%%%%%%%%%%%%%%%%%%%%%%%%%%%%

\appendix
\section{The Bundle Metric Action}\label{bundlesec}

In section \ref{actionsec}, we described the classical action for a particle moving on the configuration space  $\M_n$, in the presence of a background connection. There is a different, albeit ultimately equivalent, description of the physics that it is useful to have in mind.

\para
Given any principal bundle with connection over a Riemannian manifold, the metric on the base buddies up with the connection on the bundle to define a metric on the whole bundle.  Globally, this metric is
\be
  \tilde{g}(X,Y) = \pi^* g(X,Y) + \alpha \omega(X)\, \omega(Y)\ .\nn\ee
 Here the free variable $\alpha\in \mathbb{R}$ reflects our one-parameter choice of metric on the fibre. In the local coordinates defined in section \ref{actionsec}, we have
\be ds^2 = \tilde{g}_{AB}dX^A dX^B = g_{ab} dx^a dx^b + \alpha \,(d\sigma + A_a dx^a)^2\ .\label{bmetric}
\ee
This is the kind of metric that is familiar from Kaluza-Klein theory. We then consider the quadratic action
\be  S'[\bar{\gamma}] = \int_{\bar{\gamma}} dt\ \frac{1}{2}g_{AB} \dot{X}^A \dot{X}^B	= \int_{\bar{\gamma}} dt\  \frac{1}{2}\left(g_{ab}(x) \dot{x}^a \dot{x}^b + \alpha (\dot{\sigma} + A_a \dot{x}^a)^2\right) \ .\label{bundlemetric}\ee
This differs from the action \eqn{better} because it has a quadratic, as opposed to linear,  kinetic term for the fibre coordinate $\sigma$. At first glance that seems very different: the actions \eqn{better} and \eqn{bundlemetric} describe systems with different numbers of degrees of freedom. The relationship between them arises when we realise that the latter action has an additional conserved charge
\be Q= \frac{\partial \mathcal{L}}{\partial \dot{\sigma}} = \alpha(\dot{\sigma}+ A_a \dot{x}^a) \ .\ee
For a fixed value of $Q$, the equations of motion of \eqn{bundlemetric} coincides with those derived from our earlier action \ref{better}, provided that we identify $Q=k$. The extra degree of freedom of \eqn{bundlemetric} is simply that it includes solutions with different values of $Q$. Classically, we can have $Q\in \R$ but, quantum mechanically, we must have $Q\in \Z$. 

%Now we again compute conjugate momenta. We have
%\begin{align}
%  p_a 			&:= \frac{\partial \mathcal{L}}{\partial \dot{x}^a} = g_{ab} \dot{x}^b + 2\alpha (\dot{\sigma} + A_b \dot{x}^b)  A_a &	& \hspace{-30mm}= 	 g_{ab} \dot{x}^b + q A_a\nn\\
%  p_0 			&:= \frac{\partial \mathcal{L}}{\partial \dot{\sigma}} =  2\alpha (\dot{\sigma} + A_b \dot{x}^b) 		%&&\hspace{-30mm}= q
%\end{align}
%Thus, if we identify $q=k$, we get one we got from the other action!

\para
The action (\ref{bundlemetric}) makes the $SU(2)$ action described in section \ref{rotationsec} particularly transparent: the connection is $SU(2)$ invariant only if the metric \eqn{bmetric} has an $SU(2)$ isometry.

\section{Proof of Bundle Construction}\label{multi-particle bundle appendix}

We would like to show that the bundles $U(1) \hookrightarrow P_n[w]\to \M_n$, defined for each set of $\frac{1}{2}n(n-1)$ integers $w_{ij}=-w_{ji}$ by the pullback $P_n[w] = \phi^* \widehat{P}_n[w]$, provide a full and minimal set of representatives of the isomorphism classes of $U(1)$ principal bundles over $\M_n$. That is, we claim that each of the $P_n[w]$ is topologically inequivalent from every other, and furthermore that every $U(1)$ principal bundle over $\M_n$ is isomorphic to some $P_n[w]$.

\para
We know that the bundles $U(1)\hookrightarrow \widehat{P}_n[w] \to \widehat{\M}_n$ are topologically-inequivalent and span the isomorphism classes of bundles over $\widehat{\M}_n$. By the naturality of the second Chern class, it is therefore sufficient to show that the map
\begin{align}
  \phi^*: H^2(\widehat{\M}_n,\Z) \to H^2(\M_n,\Z)\ ,
\end{align}
is an isomorphism. Since $H_1(\M_n)=0$, by the universal coefficient theorem we have $H^2(\M_n,\Z)\cong \text{Hom} (H_2(\M_n,\Z),\Z)$. Similarly for $\wM_n$. Thus, it is sufficient to show that
\begin{align}
  \phi_* : H_2(\M_n,\Z) \to H_2(\wM_n,\Z)\ ,
\end{align}
is an isomorphism. Next, since both $\M_2$ and $\wM_2$ have trivial fundamental group, for both spaces the Hurewicz map is an isomorphism, and we thus have $H_2(\M_n,\Z)\cong \pi_2(\M_n)$ and $H_2(\wM_n,\Z)\cong \pi_2(\wM_n)$. Thus, we have reduced the problem to showing that
\begin{align}
  \phi_* : \pi_2(\M_n) \to \pi_2(\wM_n)\ ,
  \label{pi 2 iso}
\end{align}
is an isomorphism. Let us see why this is true. $\pi_2(\M_n)$ is a free Abelian group generated by elements $\rho_{ij}$ for $i<j$, defined by moving the $i^\text{th}$ point around a small $S^2$ enclosing only the $j^\text{th}$ point; a proof of this fact can be found for instance in Sec II of \cite{fadell}. Conversely, $\pi_2(\wM_2)$ is a free Abelian group generated by elements $\sigma_{ij}$ for $i<j$, defined by moving $\r_{ij}$ in an $S^2$ about the origin in its $\R^3\setminus \{\0\}$ while holding all other coordinates fixed. Observe then that $\sigma_{ij}$ is homotopic to $\phi\circ \rho_{ij}$, and thus $\phi_* \rho_{ij} = \sigma_{ij}$. We have therefore established that (\ref{pi 2 iso}) is indeed an isomorphism.

\para
Note, our proof provides an independent derivation of the result \cite{dmann}
\begin{align}
  H^2(\M_n,\Z) \cong H^2(\wM,\Z) \cong \Z^{\frac{1}{2}n(n-1)}\ .
\end{align}

\newpage 
\bibliography{dyonhilbert.bib}
\bibliographystyle{JHEP}

\end{document}